\documentclass[aip,jcp,reprint]{revtex4-1}
\usepackage{bm}
\usepackage{mathtools}
\usepackage{braket}
\usepackage{amsmath}
\usepackage{url}
\usepackage{hyperref}
\usepackage[usenames,dvipsnames,table]{xcolor}
\renewcommand{\thispagestyle}[1]{}

\begin{document}
\title{Communication: Charge-Population Based Dispersion Interactions for Molecules and Materials}
\author{Martin St\"ohr}
\affiliation{Department of Chemistry, Yale University, New Haven CT 06520, United States}
\affiliation{Department Chemie, Technische Universit{\"a}t M{\"u}nchen, Lichtenbergstr. 4, D-85748 Garching, Germany}
\author{Georg S. Michelitsch}
\affiliation{Department Chemie, Technische Universit{\"a}t M{\"u}nchen, Lichtenbergstr. 4, D-85748 Garching, Germany}
\author{John C. Tully}
\affiliation{Department of Chemistry, Yale University, New Haven CT 06520, United States}
\author{Karsten Reuter}
\affiliation{Department Chemie, Technische Universit{\"a}t M{\"u}nchen, Lichtenbergstr. 4, D-85748 Garching, Germany}
\author{Reinhard J. Maurer}
\email[]{reinhard.maurer@yale.edu}
\affiliation{Department of Chemistry, Yale University, New Haven CT 06520, United States}
\date{\today}

\begin{abstract}
We introduce a system-independent method to derive effective atomic C$_6$ coefficients and polarizabilities in molecules and materials purely from charge population analysis. This enables the use of dispersion-correction schemes in electronic structure calculations without recourse to electron-density partitioning schemes and expands their applicability to semi-empirical methods and tight-binding Hamiltonians. 
We show that the accuracy of our method is \emph{en par} with established electron-density partitioning based approaches in describing intermolecular C$_6$ coefficients as well as dispersion energies of weakly bound molecular dimers, organic crystals, and supramolecular complexes. We showcase the utility of our approach by incorporation of the recently developed many-body dispersion (MBD) method [Tkatchenko \emph{et al.}, Phys. Rev. Lett. \textbf{108}, 236402 (2012)] into the semi-empirical Density Functional Tight-Binding (DFTB) method and propose the latter as a viable technique to study hybrid organic-inorganic interfaces.
\end{abstract}

% \pacs{71.15.Qe,31.15.B-,31.15.E-,73.20.Hb}

% \keywords{dispersion interactions, many-body dispersion, density-functional tight-binding, intermolecular interactions}

\maketitle 

Long-range correlations such as dispersion interactions play an important role in the molecular structure and reaction dynamics of many materials and molecules. Many computationally efficient \emph{ab initio} electronic structure approaches, such as current approximations to Density-Functional Theory (DFT), neglect long-range dispersion interactions by construction. As a result of recent method development efforts, a number of different dispersion-inclusive \emph{ab initio} approaches have been devised, such as DFT-D3~\cite{D3} and DFT+vdW(TS)~\cite{TS-PRL-2009}, van der Waals functionals~\cite{Dion2004,Klimes2011} (vdW-DF), or the recent many-body dispersion method~\cite{MBD,rsSCSMBD} (DFT+MBD). Some of these methods provide an accurate electronic structure description of intermolecular interactions in molecular dimers~\cite{D3}, organic crystals~\cite{Reilly2013,MBDaspirin,Brandenburg2014}, hybrid organic-inorganic interfaces~\cite{Mercurio2013,MBDHIOS}, and supramolecular complexes~\cite{MBDS12L,Brandenburg2014}.

Notwithstanding recent advances in extending accurate electronic structure methods to the solid state~\cite{Michaelides2015}, specifically in the context of nanostructured materials and complex interfaces a need exists for more efficient methods with reduced scaling properties and reliable account of long-range interactions. At the cost of reduced transferability such approaches allow to address structural changes and chemical reactions at longer length and time scales. One such class of methods are semi-empirical methods and model Hamiltonians. Contrary to molecular mechanics or force fields methods that completely eliminate the explicit description of electronic structure, semi-empirical methods such as the wavefunction based Neglect-of-Diatomic-Differential-Overlap~\cite{NDDO} (NDDO) methods as represented by AM1~\cite{AM1} and PM3~\cite{PM3}, or the DFT-based FIREBALL method~\cite{Sankey1989,Lewis2001} and Density-Functional-based Tight Binding method~\cite{DFTB, SCC-DFTB} (DFTB) retain a parametrized minimal basis representation of the electronic Hamiltonian. As a result they provide access to electronic~\cite{Todorov2001}, optical~\cite{Niehaus2001}, and magnetic~\cite{Koehler2007} properties of materials.

Often derived from mean-field methods such as Hartree-Fock or semi-local DFT, these effective methods unfortunately suffer from the same intrinsic neglect of long-range interactions. Just as with semi-local DFT, they may thus in principle be coupled with semi-empirical pairwise dispersion correction approaches such as first put forward by Grimme \emph{et al.} (DFT-D)~\cite{Grimme2004}. This has e.g. been extensively done in the context of DFTB~\cite{McNamara2007,Rapacioli2009,Petraglia2015}. In such schemes dispersion corrections are incorporated by addition of the leading terms of the dipolar expansion that captures the dynamical charge fluctuations of atoms in molecules and materials
\begin{equation}\label{eq:dispcorr}
E_{\mathrm{disp}} = -\sum_{A<B}f_{\mathrm{damp}}\left(R_{AB},R_A,R_B\right)\frac{C_{6}^{AB}}{R_{AB}^6} \quad ,
\end{equation}
where $f_{\mathrm{damp}}$ is a damping function limiting the correction to distances $R_{AB}$ beyond effective van-der-Waals radii $R_A$ and $R_B$, and $C_{6}^{AB}$ correspond to the interatomic C$_6$ dispersion coefficients. All these parameters are thereby precalculated and tabulated.

In recent years, a number of approaches has been derived that provide a more profound connection between DFT and dispersion interactions by deriving dispersion coefficients from coordination numbers~\cite{D3}, the electron density~\cite{TS-PRL-2009}, the exchange-hole dipole moment~\cite{BeckeJohnson}, Wannier functions~\cite{Silvestrelli2008}, or by directly modelling a non-local density functional~\cite{Berland2015}. The above methods account for the dependence of dispersion interactions on the real-space distribution of the electron-density or the wavefunctions. The corresponding numerical schemes represent additional steps in electronic structure simulations and, for semi-empirical methods, may provide severe computational bottlenecks, notwithstanding the crude real-space representation of molecular properties in a minimal basis.

In this work we provide a connection between a given Hamiltonian in a local basis representation and atom-wise dispersion coefficients. This eliminates the recourse to the electron density in real-space, and thereby allows for an efficient coupling of advanced dispersion-correction schemes with Density Functionals and semi-empirical methods such as DFTB. Our method is based on charge population analysis (CPA) and, in contrast to other approaches~\cite{McNamara2007,Rapacioli2009,Brandenburg2014,Petraglia2015}, does not introduce additional system-dependent parametrization. It has proven insensitive to the underlying basis set representation for both tested cases of DFT and DFTB. Validation on a large number of intermolecular C$_6$ coefficients and standardized  benchmark sets for intermolecular interactions shows that our charge-population based scheme coupled with DFT yields highly accurate intermolecular C$_6$ coefficients and interaction energies when compared to experiment and high-level reference data. Our scheme is therefore \emph{en par} in accuracy with the atoms-in-molecules density partitioning-based vdW(TS) of Tkatchenko and Scheffler for a wide range of molecular systems.

The DFT+vdW(TS), or in short DFT+TS, scheme~\cite{TS-PRL-2009} represents a particularly simple and accurate method to derive dispersion interactions directly from the electron density. The dispersion interaction as given by eq.~\ref{eq:dispcorr} is defined via effective atom-wise dispersion parameters such as static atomic polarizabilities $\alpha^0_A$, C$_{6}^{AA}$ coefficients, and van-der-Waals radii R$_{A}$
\begin{equation}\label{eq:interatomic-C6}
C_{6}^{AB} = \frac{2\,C_{6}^{AA} C_{6}^{BB}}{\frac{\alpha_B^0}{\alpha_A^0} C_{6}^{AA} + \frac{\alpha_A^0}{\alpha_B^0} C_{6}^{BB}}.
\end{equation}
The effective atomic parameters for an atom in a molecule that enter eq.~\ref{eq:dispcorr} are related to accurate free atom reference data~\cite{Chu2004} by the change in atomic polarizability due to the chemical environment in which the atom is embedded. Exploiting the linear correlation of polarizability and effective atomic volume~\cite{alpha2V} the parameters are obtained as a function of the volume ratio between the free atom ($V_A^{\rm free}$) and the atom in the environment ($V_A$)
\begin{equation}\label{eq:C6-HA}
\frac{C_{6}^{AA}}{C_{6}^{AA,\mathrm{free}}} \approx \left(\frac{\alpha_A}{\alpha_A^{\mathrm{free}}}\right)^2 \approx \left(\frac{V_A}{V_A^{\mathrm{free}}}\right)^2 \quad . 
\end{equation}
The atomic volumes are thereby calculated using the Hirshfeld atoms-in-molecules density partitioning scheme~\cite{Hirshfeld1977}.

The favourable scaling properties of DFTB and similar methods stem from removal of time-consuming components such as the explicit construction of the electron density and the evaluation of multi-center integrals. This leaves the parametrized Hamiltonian in a minimal atomic orbital basis set representation. However, with no direct access to the electron density, the vdw(TS) scheme can not be applied. Inspired by a recent dispersion correction approach for force fields based on tessellation of an artificial electron density~\cite{FF+MBD} we therefore attempted to directly reconstruct the electron density in DFTB from the confined atomic orbitals used to parametrize the electronic DFTB Hamiltonian and DFTB orbital occupations (see supplemental material for details\cite{supplemental}). Unfortunately, the resulting DFTB+TS dispersion coefficients suffer from the poor density representation and show significant deviations from the DFT-based scheme and accurate reference data for a number of benchmark systems. More importantly, the results also strongly depend on the choice of confinement of the free atom basis functions employed in the DFTB scheme.

\begin{figure}
  \centering
  \includegraphics[width=3.3in,natwidth=238,natheight=223]{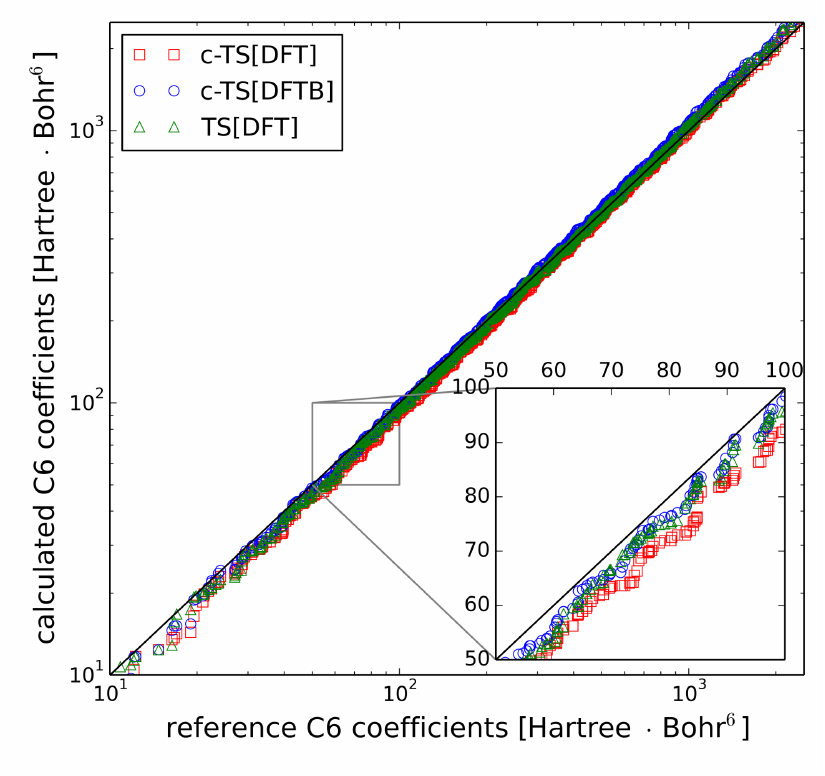}
  \caption{Comparison of interatomic C$_{6}$ coefficients as obtained by c-TS in combination with DFT-PBE (red squares) and DFTB (blue circles) against accurate reference values derived from dipole oscillator strength distributions (black line). Additionally, the corresponding values for the original TS scheme are included (green triangles). Inset: The larger systematic deviations in the region of 50--100~Ha$\cdot$Bohr$^{6}$ are the main reason for the slightly increased overall MARE in c-TS[DFT] compared to TS[DFT].}
  \label{Fig:C6-correlations}
\end{figure}

This leaves us with the need to identify an alternative relation between electronic structure and atomic polarizability. While the static atomic polarizability is directly proportional to the atomic volume, it is also indirectly proportional to the chemical hardness or the degree of hybridization~\cite{alpha2V, Miller1979, No1993}. One possible measure for the degree of hybridization when using a local atomic-orbital basis set $\ket{\psi_a}=\sum_i c_i^a \cdot\ket{\phi_i}$ is the atom-projected trace of the density matrix~\cite{Davidson1967}
\begin{equation}
 h_A = \sum_a f_a \sum_{i\in A} \left| c_i^a \right|^2,
\end{equation}
with $f_a$ the molecular orbital occupation of state $a$, and $c_i^a$ the associated wavefunction coefficient corresponding to basis function $i$ located at atom $A$. This measure thus accounts for the hybridization-induced charge transfer and effective volume change due to interaction with other atoms (see supplemental material for more details~\cite{supplemental}). It corresponds to the on-site contribution to Mulliken populations~\cite{MOOP}, which is equal to the atomic charge $Z_A$ in the case of a free atom. We therefore propose to approximate the change of polarizability of an atom in a molecule or a solid as follows
\begin{equation}\label{eq:C6-ONOP}
\frac{C_{6}^{AA}}{C_{6}^{AA,\mathrm{free}}} \approx \left(\frac{\alpha_A}{\alpha_A^{\mathrm{free}}}\right)^2 \approx \left(\frac{h_A}{Z_A}\right)^2 \quad .
\end{equation}
This CPA approach yields the correct limit for free neutral atoms and effectively accounts for bond formation and coordination. Free atom values of $\alpha^{\mathrm{free}}_A$, C$_{6}^{AA,\mathrm{free}}$, $R_{\mathrm{A}}^{\mathrm{free}}$ are correspondingly rescaled~\cite{TS-PRL-2009} and enter equation~\ref{eq:interatomic-C6}. The CPA can be employed in electron-density based dispersion correction approaches such as TS and MBD and the resulting schemes will henceforth be referred to with the prefix c, as for instance c-TS or c-MBD.

We assess the accuracy of dispersion interactions derived \emph{via} CPA by calculating a set of intermolecular C$_6$ coefficients of 817 complexes proposed as benchmark by Meath and co-workers\cite{Meath1,Meath2} on the basis of experimentally derived dipole oscillator strength distributions. Our method yields intermolecular $C_6$ coefficients with a Mean Absolute Relative Error (MARE) of 7.5\% against experiment when based on DFT-PBE states and 6.8\% when based on DFTB states, cf. Fig.~\ref{Fig:C6-correlations}. The MARE for the original density-partitioning approach of Tkatchenko and Scheffler~\cite{TS-PRL-2009} is 5.4\% in DFT-PBE, whereas density partitioning on the basis of an artificially constructed DFTB density yields 23.9\% error. The latter clearly shows the limitations of the original TS approach in combination with DFTB. Further, it is noteworthy that the maximum relative deviation in c-TS[DFT] is 29.6\% (for Li$\cdots$SiH$_4$), whereas it is 42\% (H$_2$ dimer) in the original TS[DFT] scheme. Principal component analysis reveals a similar linear correlation between calculation and reference for all approaches. The inset of Fig.~\ref{Fig:C6-correlations}, however, shows, that the higher overall relative error in c-TS[DFT] dominantly stems from a slight systematic underestimation, especially for small values of C$_6$ below 100~Ha$\cdot$Bohr$^{6}$.

With this encouraging result we proceed to incorporate our approach into different dispersion-corrected DFT methods and study realistic benchmark systems. We do this for the pairwise-additive TS scheme~\cite{TS-PRL-2009} and the many-body MBD scheme~\cite{MBD}. Both depend on a given set of atom-wise dispersion coefficients as a starting point. In the case of TS, an energy as given in eq.~\ref{eq:dispcorr} is evaluated. Throughout this work, we do not adjust or modify the damping function parameters of Tkatchenko and Scheffler~\cite{TS-PRL-2009} and simply apply a damping function as optimized for the PBE functional~\cite{PBE}. In the MBD case the dispersion parameters enter an interacting set of atom-centered quantum harmonic oscillators, which define a coupled fluctuating dipole model~\cite{CFDM1, CFDM2} to capture the non-additive many-body vdW interactions. All DFT calculations below were performed using the FHI-aims all-electron DFT code~\cite{AIMS}, with the semi-local PBE functional~\cite{PBE}. {SCC-DFTB}~\cite{DFTB-beginners} calculations have been carried out using the DFTB+ code~\cite{DFTB+} with recent mio-1.1~\cite{Elstner98} parameters. In both cases we extract the wavefunction coefficients and carry out dispersion calculations using a modified version of the Atomic Simulation Environment (ASE)~\cite{ASE}, which interfaces to standalone implementations of the TS~\cite{McNellis2009} and MBD~\cite{MBDcalc} methods.

\begin{center}
\begin{table}
\begin{tabular}{lp{0.08\textwidth}p{0.08\textwidth}p{0.08\textwidth}p{0.08\textwidth}} \hline
        & \multicolumn{1}{p{0.08\textwidth}}{S66x8$^{(a)}$} & \multicolumn{1}{p{0.08\textwidth}}{S22$^{(a)}$} & \multicolumn{1}{p{0.08\textwidth}}{X23$^{(b)}$} & \multicolumn{1}{p{0.08\textwidth}}{\textbf{Overall}}\tabularnewline\hline \noalign{\vskip 0.3ex}
PBE               & 1.55 & 2.61 & 11.95 & \textbf{5.37} \tabularnewline
PBE+TS            & 0.42 & 0.32 &  3.25 & \textbf{1.33} \tabularnewline
PBE+c-TS          & 0.35 & 0.30 &  1.27 & \textbf{0.64} \tabularnewline
PBE+MBD           & 0.32 & 0.48 &  1.11 & \textbf{0.64} \tabularnewline
PBE+c-MBD         & 0.32 & 0.60 &  1.94 & \textbf{0.95} \tabularnewline
DFTB              & 2.31 & 3.53 & 12.87 & \textbf{6.23} \tabularnewline
DFTB+c-TS         & 1.20 & 1.55 &  2.86 & \textbf{1.87} \tabularnewline
DFTB+c-MBD        & 1.13 & 1.27 &  2.54 & \textbf{1.64} \tabularnewline[0.0ex] \hline \noalign{\vskip 0.3ex}
\end{tabular}
\caption{Mean absolute deviation (MAD) in binding energies of three benchmark sets of molecular dimers and organic crystals calculated with DFT-PBE and DFTB with and without dispersion correction using the original electron density based schemes (TS and MBD) and using the CPA (c-TS and c-MBD). Results are given in kcal/mol with respect to $^{(a)}$interaction energies as obtained by CCSD(T)/CBS and $^{(b)}$experimental lattice energies.}
\label{Tab:MAD-IAE}
\end{table}
\end{center}

For validation of the CPA method we employ established benchmark systems of intermolecular interaction energies for molecular dimers in equilibrium (S22~\cite{S22}) and along dissociation curves (S66x8~\cite{S66x8}), as well as the lattice energies of 23 different organic crystals (X23)~\cite{Otero-de-la-Roza2012,Reilly2013a,Reilly2013}, see Table \ref{Tab:MAD-IAE}. Despite a 2\% larger error in intermolecular C$_6$ coefficients as  compared to the original scheme (PBE+TS), the CPA method combined with TS pairwise dispersion (PBE+c-TS) slightly improves the description of interaction energies. The original TS scheme is known to slightly overestimate polarizabilities and dispersion interactions~\cite{TS-PRL-2009,MBD}. In contrast, for c-TS we find a slight underestimation, especially when using the semi-local DFT-PBE functional, cf. the inset of Fig.~\ref{Fig:C6-correlations}. As we only consider energy differences, final interaction energies may benefit from error cancellation. In the case of PBE+c-MBD, mean absolute deviations (MAD) are equal or minimally increased compared to the original scheme. The considerably increased absolute deviations of all methods for the X23 set arise from significantly higher interaction energies with a mean of 20.4~kcal/mol. Especially for organic crystals, many-body interactions are important, yielding mean absolute relative errors (MARE) of 6.2\% and 9.8\% for PBE+MBD and PBE+c-MBD, respectively. In this specific case the underestimation of dispersion parameters in the CPA scheme when based on DFT-PBE states simultaneously improves the performance of c-TS due to error cancellation and appears to slightly impair the c-MBD scheme. Nonetheless, our approach, eq.~\ref{eq:C6-ONOP}, yields overall comparable results to the original scheme and therefore represents a viable alternative that additionally eliminates the need for electron density partitioning. 

%DFTB
Switching from DFT-PBE to the semi-empirical method DFTB, MAREs are almost consistently increased by about 10\%. This is expected from the more approximate electronic structure description at this level. Nevertheless, accounting for vdW interactions via c-TS and c-MBD drastically improves the description of intermolecular complexes and organic crystals in comparison to plain DFTB. Overall deviations are decreased from 6.2~kcal/mol (68.7\% MARE) down to 1.6~kcal/mol (23.2\% MARE) in DFTB+c-MBD. The still sizable MARE of 23.2\% may be further reduced by adaptation of the damping function of TS or the range-separation parameter of MBD to the DFTB level of description. On the other hand, the approximations made in DFTB do not produce cumulative deviations as the relative error decreases with increasing interaction energies from S66x8 to S22 and X23 (see supplemental material\cite{supplemental}). This further encourages the use of dispersion-corrected DFTB or other semi-empirical approaches for the description of extended systems.

\begin{figure}
  \centering
  \includegraphics[width=3.3in,natwidth=244,natheight=176]{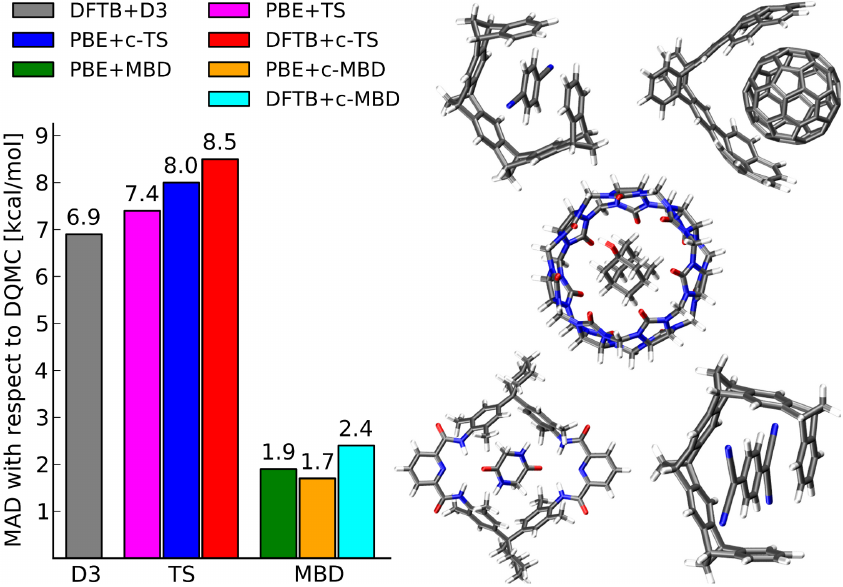}
  \caption{Left: MADs in binding energies (in kcal/mol) for a selected subset of S12L complexes as obtained by different dispersion corrected approaches with respect to DQMC calculations. right: Graphical depiction of the S12L subset considered in this work. H (white), C (black), N (blue), O (red).}
  \label{Fig:S12L}
\end{figure}
%%S12L
A particular benefit of our approach is the ability to efficiently incorporate many-body dispersion into DFT and semi-empirical methods. We further exemplify this with a set of five supramolecular complexes (a subset of the S12L benchmark set~\cite{S12L}, see Fig. \ref{Fig:S12L}) for which pairwise-additive approaches tend to overestimate binding energies, while PBE+MBD is known to perform well~\cite{MBDS12L,FF+MBD}. In reference to accurate diffusion quantum Monte-Carlo results~\cite{MBDS12L} (DQMC), our CPA method in combination with PBE and MBD (PBE+c-MBD) yields an MAD of 1.7~kcal/mol (6.0\% MARE). This is almost identical to the performance of the original PBE+MBD scheme with 1.9~kcal/mol (8.1\% MARE) for this subset. Remarkably, this level of accuracy carries over to the DFTB+c-MBD level with only slightly larger deviations of 2.4~kcal/mol (10.2\% MARE). For comparison, SCC-DFTB in conjunction with D3 including three-body interactions within the Axilrod-Teller-Muto formalism~\cite{ATM1,ATM2} yields 6.9~kcal/mol (28.5\% MARE). 

%CONCLUSION
In conclusion we presented a computationally efficient and stable approach to extract atom-wise dispersion parameters solely from the density-matrix. On-site Mulliken charges capture the trends in hybridization and effective volume that renormalize free atom dispersion coefficients for an atom embedded in a molecule or material. In conjunction with DFT, this approach eliminates the need for density partitioning and promises a considerably simplified definition of analytical forces~\cite{MBDFanalyt} when compared to density partitioning schemes. In the case of DFTB and semi-empirical methods in general it enables a parameter-free connection between atomic reference data and a parametrized Hamiltonian. At both levels of theory we find accurate intermolecular C$_6$ coefficients, intermolecular binding energies, and lattice energies for a large variety of chemical systems -- including organic dimers in (non-)equilibrium configurations, organic crystals, and supramolecular complexes. As shown for the example of supramolecular guest-host systems, DFTB in combination with many-body dispersion can yield a reliable description of stacked and intercalated complexes as they appear in porous metal-organic frameworks and hybrid organic-inorganic interfaces. A recent test study on bisphenol A aggregates adsorbed at a Ag(111) surface strongly supports this assertion~\cite{Lloyd2016}. Pending an in-depth analysis of the validity of eq.~\ref{eq:C6-ONOP} and its possible limitations in the context of inorganic and metallic materials, we suggest this method as an efficient route towards a large-scale electronic structure description of hybrid materials and complex interfaces.

%ACKNOWLEDGEMENTS
MS acknowledges financial support by the PROMOS program of the DAAD during his
stay at Yale University. RJM and JCT acknowledge funding from DoE - Basic Energy
Sciences grant no. DE-FG02-05ER15677. GSM acknowledges the support of the
Technische Universit\"at M\"unchen - Institute for Advanced Study, funded by the
German Excellence Initiative. The authors thank T. Bereau and A. Tkatchenko for
fruitful discussions and for supplying detailed information about systems
contained in the X23 and S12L benchmark sets.

%%%%%%%%%%%%%%%%%%%%%%%%%%

% \bibliography{references}


\end{document}